# Karma-OTİ Sistemler için Silinti Düzelten Reed Solomon Kodların FPGA Gerçeklemesi

# FPGA Impementation of Erasure-Only Reed Solomon Decoders for Hybrid-ARQ Systems

Cansu Şen, Soner Yeşil, Ertuğrul Kolağasıoğlu
Haberleşme ve Bilgi Teknolojileri, Mühendislik Grup Başkanlığı, Aselsan A.Ş., Ankara, Türkiye
{cansusen, syesil, ekolagasioglu}@aselsan.com.tr

*Özetçe* — Bu bildiride, Karma Otomatik Tekrar İsteği (OTİ) yöntemlerinde, Gönderim Yönünde Hata Düzeltimi (GYHD) birimi olarak Reed-Solomon (RS) kodların kullanımı anlatılmaktadır. Bu amaca uygun olacak şekilde seçilen yüksek sembol uzunluğuna sahip (örneğin GF($2^{32}$)) ve sadece silinti düzeltebilen parametrik ve esnek RS kod çözücülerin FPGA tasarımı detaylı olarak sunulmaktadır. Tek bir saat çevriminde sonuç veren GF($2^m$) çarpma birimlerine dayalı bu sistemin kod çözme hızı ve kaynak tüketimi bu birimlerin sayısı ile doğru orantılı olarak artmaktadır. Belirli bir kaynak kullanımı için kod çözme hızı, düzeltilecek silinti sayısının karesiyle azalmaktadır. Xilinx Zynq7020 SoC ailesinde elde edilen gerçekleme sonuçlarına göre, GF($2^{32}$)'de 64 silinti düzelten bir RS kod çözücü 1641-LUT, 188-FF kullanmakta ve 15Mb/s hızında kod çözmektedir. Aynı RS kod için, 8 adet paralel çarpma birimiyle kod çözme hızı 100Mb/s ve kaynak kullanımı 6128-LUT, 628-FF olarak değişmektedir.

*Anahtar Kelimeler* — *gönderim yönünde hata düzeltimi(GYHD), silinti düzelten Reed Solomon kod çözücü, FPGA, karma-OTİ*

*Abstract* — This paper presents the usage of the Reed Solomon Codes as the Forward Error Correction (FEC) unit of the Hybrid Automatic Repeat Request (ARQ) methods. Parametric and flexible FPGA implementation details of such Erasure-Only RS decoders with high symbol lengths (e.g. GF($2^{32}$)) have been presented. The design is based on the GF($2^m$) multiplier logic core operating at a single clock cycle, where the resource utilization and throughput are both directly proportional to the number of these cores. For a fixed implementation, the throughput inversely decreases with the number of erasures to be corrected. Implementation in Zynq7020 SoC device of an example GF($2^{32}$)-RS Decoder capable of correcting 64-erasures with a single multiplier resulted in 1641-LUTs and 188-FFs achieving 15Mbps, whereas the design with 8 multipliers resulted in 6128-LUTs and 628-FFs achieving 100Mbps.

*Keywords* — *forward error correction(FEC), erasure only reed solomon decoder, FPGA, hybrid ARQ*

## I. GİRİŞ

Çoğa-gönderim, bilgiyi tek göndericiden birden çok alıcıya dağıtmak için kullanılan etkili bir yoldur. Bu yöntemin temel amacı, kablosuz ağlardaki trafik yükünün azaltılması ve spektrumun daha verimli kullanılmasıdır. Kablosuz kanallarında sıkça karşılaşılan sönümleme, girişim ve ağ katmanındaki tıkanmalar gibi bozucu etkiler nedeniyle meydana gelen paket kayıpları, çoğa-gönderime dayalı bir sistemin hem tasarımı hem de sonrasında idamesi için problem oluşturmakta ve bu nedenle beraberinde bir takım önlemlerin kullanılmasını zorunlu kılmaktadır [1][2]. Verinin çoğa-gönderim yoluyla kablosuz kanallar gibi güvenilir olmayan bir ortam üzerinden iletiminde paket kayıplarının önüne geçilmesi için genellikle Gönderim Yönünde Hata Düzeltimi (GYHD) ve Otomatik Tekrar İsteği (OTİ) yöntemlerinin birlikte kullanıldığı Karma-OTİ yöntemleri yaygındır [3]. Karma-OTİ sistemlerde GYHD vasıtasıyla hatalı paketler alma tarafında düzeltilerek yeniden iletim miktarları azaltılmakta ve böylece kanal daha verimli kullanılmaktadır. GYHD'nin yetersiz kaldığı durumlarda ise paketlerin alıcı tarafından yeniden istenmesi ile sistem güvenilirliği artmaktadır. GYHD ile OTİ yöntemlerini birlikte kullanan sistemler sadece GYHD kullananlara göre daha güvenilir, sadece OTİ kullananlara göre ise daha verimli kanal kullanımına sahip olmaktadır [4].

Karma-OTİ sistemlerde GYHD yöntemleri olarak genellikle Fountain, Luby Dönüşüm ve Raptor kodları kullanılmaktadır[5]. Yoğun hata gruplarına karşı gösterdiği direnç sayesinde, özellikle veri depolama ve uzak mesafe uzay haberleşme sistemlerinde kanal kodlama amacıyla yaygın olarak yer bulmuş olan Reed Solomon (RS) Kodlar, son yıllarda Karma-OTİ sistemler için de kullanılmaya başlanmıştır[6]. RS kodların kablosuz çoğa-gönderim senaryolarında diğer kodlara kıyasla tercih edilmesinin en önemli nedeni, tekrar gönderim ve düzeltilecek silinti sayısının kesin olarak belirli olmasıdır. Bu özellikleri nedeniyle RS kodlar, gecikmelerin en aza indirgendiği ve spektrumun daha verimli kullanıldığı basit ve gürbüz ağ katmanı tasarımlarının hayata geçirilmesine olanak sağlar. Bu kodların önemli bir yitirimi ise kod çözümündeki işlem karmaşıklığının yüksek olmasıdır. Silinti sayısının karesiyle artan bu karmaşıklık, RS kodların, ağ seviyesi işlemlerin koştuğu ve çoğunlukla işlemci tabanlı ortamlarda gerçeklenmesinde engel teşkil etmektedir. RS kod gerçeklemeleri, genellikle FPGA gibi paralel işlem yeteneği olan ancak maliyeti daha yüksek ortamlarda gereksinimleri karşılamaktadır.

Bu bildiride, GF($2^m$)'de tanımlı ve sadece silinti düzelten bir RS kod çözücünün esnek ve tekrar kullanılabilir bir yapıda FPGA gerçeklemesi anlatılmaktadır. RS kodların kısa bir tanımı ile devam eden bildiri, genel kod çözme yöntemi ile birlikte sadece silinti düzeltmenin getirdiği bazı gerçeklemeye yönelik avantajların anlatılmasının ardından esnek yapıdaki FPGA gerçekleme sonuçları ve bunlara ilişkin değerlendirme ve öneriler ile son bulmaktadır.

## II. REED SOLOMON KODLAR

Galois Field ($2^m$) üzerinde ($n, k$) parametreleriyle tanımlanan bir RS kod sisteminde, $m$-bit uzunluğundaki $k$-sembol veri ve $n$-sembol kod kelimesi, sırasıyla $D(z) = d_0 + d_1 z \cdots + d_{k-1} z^{k-1}$ ve $C(z) = c_0 + c_1 z \cdots + c_{n-1} z^{n-1}$ polinomlarıyla ifade edilir. Tüm kod kelimeleri, $G(z) = \prod_{i=0}^{n-k-1}(z - \alpha^{m_0+i})$ olarak tanımlanan üreteç polinomunun tam katlarıdır. Yukarıdaki ifadede α primitiv eleman olup $m_0$ genellikle 0 ya da 1 olarak alınır. Verinin, blok olarak yer aldığı sistematik bir kod kelimesi ise $C(z) = Q(z) \times G(z) - P(z)$ olarak ifade edilmektedir. Burada $Q(z)$ herhangi bir polinom olup $P(z)$ ise verinin sonuna eklenen parite sembollerinin polinom gösterimidir. RS kodlama ile $e + 2t \le n-k$ olacak şekilde $e$ adet silinti ve $t$ adet hatalı sembol düzeltilebilir. Bu bildiri boyunca $t$=0 kabul edilerek sadece silinti düzelten bir RS kod sistemini içeren çalışmalar anlatılmaktadır.

Hatalara maruz kalan kod kelimesi, kanal çıkışında $R(z) = C(z) + E(z)$ olacak şekilde bir hata polinomu ile toplanmış olarak modellenmektedir. $e$-adet silinti içeren bir durum için hata polinomu $E(z) = Y_1 z^{i_1} + Y_2 z^{i_2} + \cdots + Y_e z^{i_e}$ şeklinde ifade edilmektedir. Silinti yerleri önceden bilinen bu sistemde kod çözücünün görevi hata değerlerini ($Y_{j\epsilon\{1,2,\ldots,e\}}$) bulmaktır [7].

## III. REED SOLOMON SİLİNTİ DÜZELTME

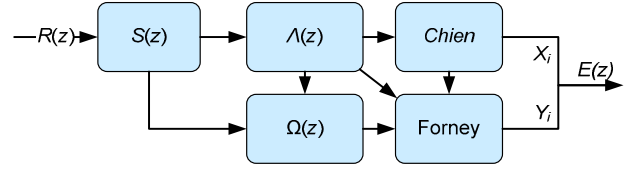

**Şekil 1**. Reed Solomon Kod Çözme Adımları

Genel yapısı **Şekil 1** de özetlenen sendrom tabanlı RS kod çözme işlemlerinde $S(z)$, $\Lambda(z)$ ve $\Omega(z)$ sırasıyla sendrom, hata-yer ve hata-değer polinomlarını ifade etmektedir. Chien Arama Algoritması bütün elemanları deneyerek $\Lambda(z)$ polinomunun köklerini bulur. $X_j$ olarak gösterilen bu köklerin GF($2^m$)'deki tersleri hata yerlerini vermektedir. Forney Birimi ise hata-yer ve hata-değer polinomlarını kullanarak hataların değerini hesaplayan bir işlemi temsil etmektedir.

Sadece silinti düzelten bir RS kod çözücünün genel bir RS kod çözücüye göre gerçekleme kazanımları bulunmaktadır. Hata yerleri önceden bilindiği için $\Lambda(z)$ polinom katsayıları kolayca hesaplanmaktadır. Kod çözme işlemindeki toplam gecikmenin önemli bir kalemini oluşturan Chien birimi kullanılmaz ve böylece yüksek kod çözme hızları elde edilebilir. Ayrıca, bu birimin kullanılmaması sadece silinti düzelten RS kodların 32-bit gibi yüksek sembol genişliklerinde tasarlanmasını pratikte mümkün kılmaktadır. $S(z) = s_0 + s_1 z \cdots + s_{n-k-1} z^{n-k-1}$ olarak gösterilen Sendrom polinomunun katsayıları, $G(z)$ köklerinin, $R(z)$ polinomundaki değerleri hesaplanarak bulunur: $s_i = R(\alpha^{m_0+i}) = C(\alpha^{m_0+i}) + E(\alpha^{m_0+i}) = E(\alpha^{m_0+i})$. Hata yer polinomu ise, $\Lambda(z) = \prod_{j=1}^{e}(1 - X_j z) = 1 + \lambda_1 z + \cdots + \lambda_e z^e$ şeklinde elde edilmektedir. Hata değerlerinin bulunması için başka bir polinom olan $\Omega(z)$'ye ihtiyaç duyulur. Bu polinom $\Lambda(z)$ ve $S(z)$ polinomlarıyla $\Lambda(z)S(z) = \Omega(z) \mod z^{(n-k)}$ anahtar bağıntısı ile bağlıdır. Son olarak, hata değerlerini bulmak için Forney formülü kullanılmaktadır (1). Paydada yer alan $\Lambda'(z)$ formal türevi göstermektedir.

$$Y_j = -\frac{z^{m_0}\Omega(z)}{z\Lambda'(z)} \Big|_{z=X_j^{-1}} \quad (1)$$

## IV. FPGA GERÇEKLEME

Bu bildiride tanımlanan sadece silinti düzelten RS kod çözücü FPGA gerçeklemesi, temel olarak tek bir saat çevriminde sonuç veren çarpma birimlerinin zamanda paylaşılmasına dayanmaktadır. GF($2^m$) üzerinde $C = AB$ şeklinde tanımlanan çarpma işlemi $C(x) = A(x)B(x) \mod P(x) = \sum_{i=0}^{m-1} c_i x^i$ polinom gösterimi ile ifade edilmektedir. $P(x)$ polinomu GF($2^m$) alan elemanlarının oluşturulduğu primitiv polinom olacak şekilde çarpma işlemi denklem (2)' deki matris çarpımı olarak gösterilmektedir[8]:

$$C = \begin{pmatrix} f_{0,0} & \cdots & f_{0,m-1} \\ \vdots & \ddots & \vdots \\ f_{m-1,0} & \cdots & f_{m-1,m-1} \end{pmatrix} \begin{pmatrix} b_0 \\ b_1 \\ \vdots \\ b_{m-1} \end{pmatrix} = ZB \quad (2)$$

Bu gösterimde i=0,1,…,m-1 olacak şekilde Z-matrisinin girişleri, $A(x) = \alpha_0 + \alpha_1 x^1 + \cdots + \alpha_{m-1} x^{m-1}$ polinom katsayılarına aşağıdaki gibi bağlıdır:

$$f_{i,j}(A) = \begin{cases} \alpha_i & j=0 \\ u[i-j]a_{i-j} + \sum_{t=0}^{j-1} q_{j-1-t,i} a_{m-1-t} & j=1,\dots,m\text{-}1 \end{cases} \quad (3)$$

Denklem (3)'te yer alan $u[n]$ birim basamak fonksiyonu olup, $q_{ij}$ değerleri $P(x)$'e aşağıdaki gibi bağlı bir Q-matrisinin girişlerini oluşturmaktadır.

$$\begin{pmatrix} x^m \\ x^{m+1} \\ \vdots \\ x^{2m-2} \end{pmatrix} = \begin{pmatrix} q_{0,0} & \cdots & q_{0,m-1} \\ \vdots & \ddots & \vdots \\ q_{m-2,0} & \cdots & q_{m-2,m-1} \end{pmatrix} \begin{pmatrix} 1 \\ x \\ \vdots \\ x^{m-1} \end{pmatrix} \bmod P(x) \quad (4)$$

$GF(2^m)$ çarpma sonucundaki her bir bit, $A(x)$, $B(x)$ girdileri ve $P(x)$ primitiv polinomlarının ikilik düzendeki katsayılarının mantıksal bir fonksiyonu olarak ifade edilmektedir. Bu fonksiyonların FPGA sentez sonucunda ortaya çıkan kaynak kullanımı ve gecikme seviyeleri $P(x)$ polinomunun Hamming ağırlığı arttıkça artmaktadır. Primitiv polinomlar bu bakış açısıyla incelendiğinde, en az kaynak kullanımı ve mantıksal gecikme seviyelerini "Trinomial" olarak adlandırılan ve Hamming ağırlığı 3 olan polinomlar vermektedir. $GF(2^{32})$ için trinomial bir primitiv polinom olmadığı bilinmektedir[9]. Ancak bu alan için çeşitli "Pentanomial" (Hamming ağırlığı=5) primitiv polinomlar mevcuttur. **Tablo 1**'de yer alan pentanomial polinomlar ile tanımlanan $GF(2^{32})$ çarpma işlemleri FPGA üzerinde gerçeklenerek kaynak kullanımları elde edilmiştir.

| $GF(2^{32})$ Primitiv Polinomlar: $P(x)$ | LUT | F7Mux | F8Mux | Slice |
|---|---|---|---|---|
| [32,25,15,7,0] | 934 | 58 | - | 258 |
| [32,28,27,1,0] | 705 | 8 | - | 200 |
| [32,16,7,2,0] | 708 | 5 | - | 198 |
| [32,7,6,2,0] | 558 | 11 | 1 | 151 |
| **[32,31,3,1,0]** | **541** | **-** | **-** | **148** |

**Tablo 1:** $P(x)$'e göre $GF(2^{32})$ çarpma işlemi kaynak kullanımı (XC7020)

Yukarıdaki kriterler sonucunda seçilen $P(x) = 1 + x^1 + x^3 + x^{31} + x^{32}$ polinomu ile oluşturulan bir RS kod sisteminde, (2)-(3) ve (4)'te anlatılan yöntemlere dayalı olarak FPGA'da gerçeklenen $GF(2^{32})$ çarpma birimi **Şekil 2**'deki yapılarda ve zamanda paylaşılarak kod çözücü algoritmalarında kullanılmaktadır.

$GF(2^m)$ çarpma ve toplama (XOR) birimleri ve gecikme elemanlarından oluşan ve basit anahtarlama yöntemleriyle birbirlerine dönüştürülebilen mantıksal İşlem Birimlerinin (İB) istenilen sayıda paralel olarak kullanılmasıyla kaynak kullanımı ve kod çözme bakımından esnek bir tasarım elde etmek mümkündür (**Şekil 3**).

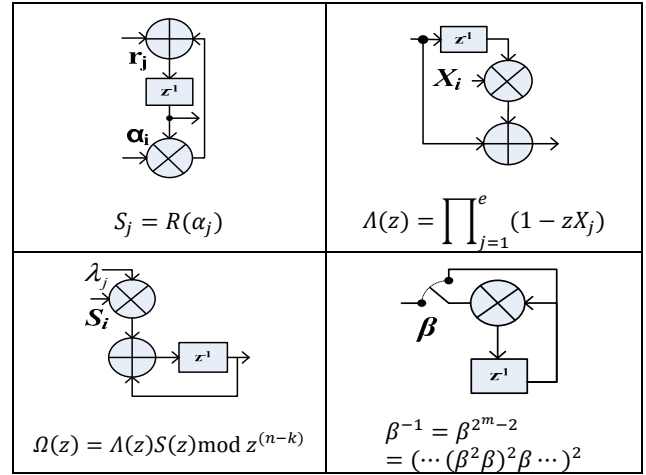

**Şekil 2**: Çarpma biriminin $S(z)$, $\Lambda(z)$, $\Omega(z)$ ve ters alma işlemleri için farklı yapılarda kullanımı.

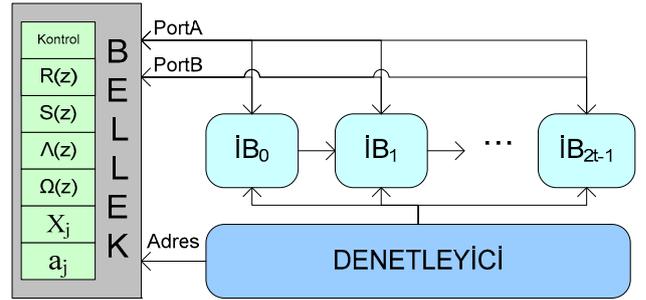

**Şekil 3:** RS Kod çözücü FPGA mimarisi

Bu yapı ile elde edilen $\Lambda(z)$ ve $\Omega(z)$ polinom katsayıları, (5)'de verilen Forney bağıntısında silinti noktalarındaki gerçek sembol değerlerini ($Y_j$) bulmak için kullanılır. Bu formülün gerçeklenmesini basitleştirmek adına $GF(2^m)$'deki çıkarma ve toplama işlemlerinin özdeş olmasının bir sonucu olarak türev almak yerine $z\Lambda'(z) = z(\lambda_1 + 2\lambda_2 z + 3\lambda_3 z^2 + \cdots) = (\lambda_1 z + \lambda_3 z^3 + \lambda_5 z^5 \dots)$ eşitliği kullanılmaktadır[7]. Ayrıca, Forney formülünü basitleştirmek için $m_0 = 0$ olarak seçilmiştir. Yüksek $m$ değerleri için $GF(2^m)$ ters alma işlemi maliyet ve gecikmelere neden olacağı için, diğer bir yöntem olarak, ters almak yerine doğrudan $X_j$ değerlerini kullanabilmek için, pay ve payda $z^{-(n-k)}$ ile çarpılarak formül (5)' teki haline getirilmiştir.

$$Y_j = \frac{z^{-(n-k)}\Omega(z)}{z^{-(n-k)}z\Lambda'(z)} \Big|_{z=X_j} \quad (5)$$

Yukarıdaki formülde pay ve payda her $X_j$ için hesaplandıktan sonra paydanın $GF(2^m)$ alanındaki tersi hesaplanıp pay ile çarpılarak silinti değerleri bulunur. Ters alma işlemi için $\beta^{-1} = \beta^{2^m-2} = (...(\beta^2\beta)^2\beta...)^2$ şeklinde özetlenen ve ($2m$-3) adet çarpma işleminin kullanıldığı bir dizi kare alma ve kendisiyle çarpma işlemleri kullanılmaktadır.

Kod çözme hızı silinti sayısının karesiyle azalmaktadır. **Şekil *4*'**te bir adet çarpma biriminin kullanıldığı bir tasarımda, farklı uzunluklarda kodlanmamış veri ve maksimum silinti sayısına göre kod çözme hızı değişimi özetlenmektedir. Yüksek sayıda silinti düzelten RS kodların kullanımında, FPGA kaynak gereksinimlerinin izin verdiği ölçüde, çarpma birimlerin sayısı arttırılarak kod çözme hızını arttırmak mümkündür. Kaynak kullanımı çarpma birimlerinin sayısı ile doğru orantılı olarak artmaktadır. **Tablo *2*'**de, silinti sayısının $e=(n-k)$ olduğu en kötü senaryodaki kod çözme işlemlerinde harcanan yaklaşık saat çevrim sayısı verilerek işlem karmaşıklığı $n$, $k$, $m$ ve $P$ değişkenleri cinsinden özetlenmiştir. **Tablo *3*'**te ise örnek bir RS(200,136) kod için aynı anda çalışan çarpma birim sayısına göre elde edilen kaynak kullanımı ve 100MHz saat hızındaki kod çözme hız değerleri gösterilmektedir.

| İşlem | Saat Çevrim Sayısı |
|---|---|
| Silinti Yerleri: $X_j$ | $N_x \sim n + 2e$ |
| Sendrom Hesaplama: $S(z)$ | $N_S \sim n(n-k)/P$ |
| Hata Yer: $\Lambda(z)$ | $N_\Lambda \sim e^2/2P$ |
| Anahtar Denklem: $\Omega(z)$ | $N_\Lambda \sim e \cdot (n-k)/2P$ |
| Forney Pay ve Payda | $N_F \sim [e^2 + (n-k) \cdot e]/P$ |
| Ters Alma | $N_{\beta^{-1}} \sim (2m) \cdot e/P$ |

**Tablo 2:** $e$-adet silinti düzeltimi için harcanan yaklaşık saat çevrim sayısı

| Kaynak-Hız Ölçütleri | P=1 | P=2 | P=4 | P=8 |
|---|---|---|---|---|
| FPGA-LUT | 1641 | 2282 | 3564 | 6128 |
| FPGA-FF | 188 | 250 | 376 | 628 |
| RAMB36 | 1 | 1 | 1 | 1 |
| Kod Çözme Hızı (Mbps@100MHz) | 14.7 | 29.1 | 57.1 | 101 |

**Tablo 3:** GF($2^{32}$)-RS(200, 136) 64-silinti düzelten kod çözücü için farklı parallel çarpma birimlerine göre kaynak ve hız bilgisi

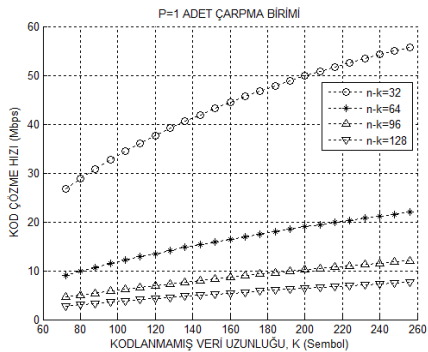

**Şekil 4:** Farklı veri uzunlukları ve maksimum silinti sayılarına göre kod çözme hızı eğrileri (fclk=100MHz)

## V. SONUÇLAR VE ÖNERİLER

Çalışmamızda GF($2^{32}$) alan üzerinde silinti düzelten bir tasarımın FPGA kaynak kullanımı ve işlem gecikme analizi yapılmıştır. Bu çalışmada, çarpma işlemi için tek bir saat çevriminde sonuç veren FPGA birimleri kullanılmıştır. Aynı anda çalışan bu birimlerin sayısına bağlı olarak, belirli bir maksimum silinti sayısına göre kod çözme hızı ve kaynak kullanım değişimi incelenmiştir.

Bu çalışma, burada detayları verilen çarpma birimlerinin yardımcı işlemci olarak kullanıldığı işlemci tabanlı bir kod çözücü ile daha esnek ve parametrik bir yapıda tekrarlanabilir. Forney formülündeki payda değerinin ters alma işlemi yüksek m değerleri için kod çözme hızını olumsuz etkilemektedir. Daha hızlı ters alma yöntemleri veya farklı matematiksel yöntemler ile bu durum iyileştirilebilir. Burada anlatılan yöntemlerden farklı olarak, matris evrimi ile veya sendroma bağlı olmayan yöntemlerle sadece silinti düzelten RS kod çözücü gerçeklemeleri mümkündür. Bu yöntemlerin incelenmesi ve buradaki sonuçlar ile karşılaştırılması da bu ve benzer çalışmaları daha iyiye götürecek bir araştırma konusu olarak değerlendirilmektedir.

### KAYNAKÇA

[1] Luby, M., Vicisano, L., Gemmell, J., Rizzo, L., Handley, M., and Crowcroft J., "The Use of Forward Error Correction (FEC) in Reliable Multicast", *RFC 3453*, December 2002.

[2] Rizzo, L., "Effective Erasure Codes for Reliable Computer Communication Protocols" *ACM SIGCOMM Computer Comm. Review*, Vol.27, No.2, pp.24-36, Apr 1997.

[3] McAuley J.A., "Reliable Broadband Communication Using a Burst Erasure Correcting Code", in *Proc. ACM SIGCOMM '90;* Sept. 1990, pp. 297–306.

[4] Lin, S. and Yu, S.P. "A Hybrid ARQ Scheme with Parity Retransmission Error Control of Satellite Channels", *IEEE Trans. on Commun.*, vol.com-30, NO. 7, July 1982.

[5] Gül, G., Adhikari, S. and Mustafin, E. "Fountain Codes, LT Codes and Raptor Codes".

[6] Lacan, J., Roca, V., Peltotalo J., Peltotalo S., "Reed-Solomon Forward Error Correction (FEC) Schemes", *RFC 5510*, April 2009.

[7] Sarwate, D.V., Shanbhag, N.,R., "High Speed Architectures for Reed Solomon Decoders", *IEEE Trans. on very large scale integ.(VLSI) sys.*, vol.9, NO. 5, October 2001.

[8] Edoardo, M., "VLSI Architectures for Computations in Galois Fields", Phd Thesis, Linköping Studies in Science and Technology Dissertation No.242, 1991.

[9] R.G. Swan. Factorization of polynomials over Finite Fields. Pacific J. Math.,12,pp.1099-1106, 1962.